# Elastic Quantum Coupling Between Free Electrons and Photons

Dingguo Zheng and Ofer Kfir[*]

*School of Electrical and Computer Engineering, the Iby and Aladar Fleischman Faculty of Engineering,*

*Tel Aviv University, Tel Aviv 69978, Israel*

**Abstract**: The quantum coupling between free-electrons and photons enables applying quantum optics techniques in electron microscopy. Here, we formulate the elastic electron-photon quantum coupling and its possible implications. Our analysis shows that when an electron traverses the field of an optical cavity, it induces a phase shift onto its confined photonic mode, which can be quantified as a refractive index of a free electron. This principle can be applied to counting electrons in a beam without changing its quantum states. The elastic scattering operator forms an electron-counting dispersive Hamiltonian for electron-photon systems within electron microscope, and it could enable quantum- and sub-shot-noise sensing and imaging at the Å-scale.

## I. INTRODUCTION

Quantum shot noise—arising from the random arrival of discrete carries—sets a fundamental limit on the sensitivity across a variety of instruments [1]. In electron microscopes, electron shot-noise ultimately constrains spatial resolution and image quality [2,3]. A major appeal in quantum metrology is the capability to suppress shot-noise fluctuations by harnessing quantum illumination [4–6].

The emergence of free electron beams with quantum statistical properties, such as heralding and entanglements [7–10], has a potential for low dose and sub-shot-noise imaging and spectroscopy in next generation electron microscopes. However, relying on a co-incident detection of a generated photon to infer the counting of one electron is accompanied by an electron energy loss, making it distinguishable from its initial state. Thus, it cannot be applied on state-preserving counting.

Studies on laser manipulation of electron beam provides possible ways for non-destructive electron detection [11–15]. Electron beams can be modulated by laser in energy space [16–20], bunched into a series of attosecond pulses [21–24] or converted into vortex beam [25,26]. In the processes of coherent light emission by free electrons, the electron passage can be detected by collecting and analyzing the generated photon's numbers and states [27–31]. The inelastic coupling strength between a photon and a free electron, $g_{\mathrm{Qu}}$, can near unity (~one photon generation for each electron) if the electron velocity matches the phase velocity of the traveling light, allowing them to exchange exact quanta of energy and momentum [32–34]. However, one possible issue with the abovementioned electron-photon interactions is that they are inelastic, so once a segment of an electron wavefunction undergoes such a process prohibits interference with other segments, such as in holography.

In elastic interaction processes free electrons acquire a phase proportional to the local intensity of a classical optical field, integrated over their trajectory. Thus, injecting lasers into the path of the beam in electron microscopes enables diffraction by standing optical lattices in the Kapitza-Dirac effect [35–37], applying a laser-based electron-phase retarder for holographic imaging [38–41], or using light to dress the electron wavefunction, either for forming the electron into arbitrary shapes [42,43], or for aberration correction [44]. However, the previous studies of elastic free-electron-photon interaction don't show detectable physical quantities to indicate electron passages.

Here, we utilize quantum optical framework for elastic electron-photon coupling that addresses the electron and photon components on an equal footing. We show where the laser-based electron-phase retardation arises from, and its back-action on the photon field. Thus, state-preserving measurement of the electron number in an electron beam is proposed. We exemplify our approach's implementation using a strong laser field. Interestingly, we show that the electron contributes an additional index-of-refraction for the light field, and find that it is birefringent, where the optical axis is along the electron velocity. The electron-induced effects are also shown for various optical quantum states. Harnessing elastic and state-preserving quantum electron-photon scattering could enable the generation of sub-Poisson multi-electron and multi-photon states, approaching the limit of electron-number states for quantum sensing down to the atomic scale.

---

[*] Contact author: kfir@tauex.tau.ac.il

We address the above points consecutively, as sections. In section II we briefly show an analytical derivation of the relevant electron-photon coupling term. Sections III and IV quantify the electron-induced photonic phase shift and reproduce the known electron phase shift induced by classical fields. Section V forms the photonic phase shift as an effective refractive index of electron beam. Section VI proposes an experimental realization, and section VII includes a discussion and outlook.

## II. QUANTUM THEORY OF ELASTIC INTERACTION

We begin by following the derivations of García de Abajo *et al.* [11,42,45], considering a propagating spin-less electron with initial energy $\varepsilon_0$ and momentum $\boldsymbol{p}_0$. The electron's relativistic parameters $\beta = v_0/c$ and $\gamma = 1/\sqrt{1-\beta^2}$ can be derived using the speed of light, $c$, and the electron velocity, $\boldsymbol{v}_0$. The vector potential operator of monochromatic light fields, $\widehat{\boldsymbol{A}}$, can be expressed as $\widehat{\boldsymbol{A}} = \boldsymbol{A}\hat{a}e^{-i\omega t} + \boldsymbol{A}^*\hat{a}^\dagger e^{i\omega t}$, where $\boldsymbol{A} = -\frac{i}{\omega}\boldsymbol{E}(\boldsymbol{r})$, $\omega$ is light frequency, $\hat{a}$ ($\hat{a}^\dagger$) is photon annihilation (creation) operator for a particular optical mode, and $\boldsymbol{E}(\boldsymbol{r})$ is the electric field associated with a single photon in that mode. The multi-electron state is described by $\hat{c}_k$ and $\hat{c}_k^\dagger$, the fermionic annihilation and creation operators, respectively, for an electron with wavevector $\boldsymbol{k}$ parallel to $\hat{z}$, the beam axis. An initial state with $N_e$ electrons can be written as $|\psi_e\rangle = \prod_{k=k_1}^{k_{N_e}} \hat{c}_k^\dagger |0\rangle_e$, where $|0\rangle_e$ is the electron vacuum state. The initial state of the system can be written as a joint electron-photon state [27], $|\Psi_i\rangle = |\psi_{ph}\rangle \otimes |\psi_e\rangle$.

The system's Hamiltonian, $\widehat{H} = \widehat{H}_0 + \widehat{H}_{int}$, includes the uncoupled term and interaction term. The latter is derived explicitly in Appendix A. However, a short outline of the derivation appears here to enable us to point to the origin of the elastic electron-photon coupling. We expand the Hamiltonian to a second order around an electron energy $\varepsilon_0$ and use the minimal coupling scheme with the mechanical momentum $\hat{p} + e\widehat{\boldsymbol{A}}$, where $e$ is the elemental charge. $\widehat{H}_0$ sums the independent electron and photon terms, $\widehat{H}_0 = \widehat{H}_{ph} + \widehat{H}_e$, where

$$\widehat{H}_{ph} = \hbar\omega\left(\hat{a}^\dagger\hat{a} + \frac{1}{2}\right), \quad (1a)$$

$$\widehat{H}_e = \sum_k \varepsilon_k \hat{c}_k^\dagger \hat{c}_k. \quad (1b)$$

For electrons the energy variance is far smaller than mean energy, so $\varepsilon_k \approx \varepsilon_0$. $\widehat{H}_{int}$ can be expressed as

$$\widehat{H}_{int} = \sum_k (\widehat{H}_1 + \widehat{H}_2 + \widehat{H}_\phi)\hat{c}_k^\dagger \hat{c}_k.$$

We separate it to terms related to phase-matching,

$$\widehat{H}_1 = e\boldsymbol{v}_0 \cdot \boldsymbol{A}\hat{a}e^{-i\omega t} + \text{c.c.} \quad (2a)$$

$$\widehat{H}_2 = \frac{e^2}{2\gamma m_e}\boldsymbol{A}^T\boldsymbol{\Gamma}\boldsymbol{A}\hat{a}^2 e^{-2i\omega t} + \text{c.c.}, \quad (2b)$$

and an elastic term independent of phase-matching,

$$\widehat{H}_\phi = \frac{e^2}{\gamma m_e}\boldsymbol{A}^\dagger\boldsymbol{\Gamma}\boldsymbol{A}\left(\hat{a}^\dagger\hat{a} + \frac{1}{2}\right). \quad (2c)$$

$m_e$ is the electron rest mass. $\boldsymbol{\Gamma}$ is a 3×3 matrix, called here the birefringence matrix. If electrons propagate along z-direction, $\boldsymbol{\Gamma}$ reduces to $\text{diag}\left(1,1,\frac{1}{\gamma^2}\right)$, and its tensor form is the origin of a birefringence effect discussed in section V.

In the interaction picture, the evolution is governed by the scattering operator, $\hat{S}$, by $|\Psi_f\rangle = \hat{S}|\Psi_i\rangle$. Solving Schrödinger equation with the interaction Hamiltonian in Eq. (2), one can find the expression for the scattering operator

$$\hat{S} = e^{i\sum_k \hat{c}_k^\dagger \hat{c}_k \chi} e^{\sum_k \hat{c}_k^\dagger \hat{c}_k \left(g_{Qu}\hat{b}\hat{a}^\dagger - g_{Qu}^*\hat{b}^\dagger\hat{a} + \frac{1}{2}g_2^*\hat{b}^{\dagger 2}\hat{a}^2 - \frac{1}{2}g_2\hat{b}^2\hat{a}^{\dagger 2} - ig_\phi\left(\hat{a}^\dagger\hat{a}+\frac{1}{2}\right)\right)} \quad (3)$$

We refer to Appendix B for a detailed derivation. Here, $\hat{b}$ and $\hat{b}^\dagger$ are the commuting electron-energy-ladder operators [27,28] in the non-recoil approximation, $g_{Qu}$ [7,8,27–29] and $g_2$ are the first and second order inelastic quantum coupling amplitudes emanating from $\widehat{H}_1$ and $\widehat{H}_2$ respectively. The elastic coupling term

$$g_\phi = \frac{e^2}{\hbar p_0}\int_{-\infty}^{+\infty} dz\, \boldsymbol{A}^\dagger\boldsymbol{\Gamma}\boldsymbol{A} \quad (4)$$

originates from $\widehat{H}_\phi$, as an integral along electron trajectory. $g_\phi$ is a real number since the Hermitian quadratic integrand $\boldsymbol{A}^\dagger\boldsymbol{\Gamma}\boldsymbol{A}$ is real. Thus, $g_\phi$ contributes pure phase-related effects to electron-photon scattering. This is the elastic coupling term we emphasize in this work. The additional global phase term, $\chi$, was investigated in detail in ref. [46]. It originates from the non-commutating nature of the interaction at different times, $[\hat{H}_{\text{int}}(t_1), \hat{H}_{\text{int}}(t_2)] \neq 0$. To the leading order of $[\hat{H}_1(t_1), \hat{H}_1(t_2)]$, one finds that

$$\chi = -\text{Im}\left\{\int d\omega \int_{-\infty}^{+\infty} dz\, g^*_{\text{Qu}}(z) \frac{d}{dz} g_{\text{Qu}}(z)\right\}.$$

Additional details on the derivation of $\chi$ shown Appendix C.

In the absence of phase matching, the terms, $g_{\text{Qu}}$ and $g_2$ vanish. However, $g_\phi \neq 0$ since it is independent of the optical phase velocity. The vacuum fluctuation phase $\chi$ is also nonzero. Therefore, the scattering operator in Eq. (3) reduces to

$$\hat{S} = e^{i\chi \hat{N}_e} e^{-ig_\phi \hat{N}_e \left(\hat{N} + \frac{1}{2}\right)}. \quad (5)$$

This photon-number dependent scattering operator is our main result. Applying the scattering operator to the initial state, we have the final state, $|\Psi_f\rangle = e^{i\chi \hat{N}_e} e^{-ig_\phi \hat{N}_e \left(\hat{N} + \frac{1}{2}\right)} |\psi_{\text{ph}}\rangle \otimes |\psi_e\rangle$. In the rest of article, the vacuum fluctuations' phase factors, $e^{i\chi \hat{N}_e}$ and $e^{-\frac{1}{2}ig_\phi \hat{N}_e}$ are omitted, since they act as a global phase, independent of any photonic excitation.

The scattering operator gives a dispersive interaction that counts the electron number $\hat{N}_e$. The form of a dispersive Hamiltonian becomes more apparent using the undisturbed photonic Hamiltonian in Eq. (1), as $\hat{S} = \exp\left(-i\frac{g_\phi}{\hbar\omega} \hat{N}_e \hat{H}_{\text{ph}}\right)$, which commute with each individual subsystem,

$$[\hat{S}, \hat{H}_e] = 0, [\hat{S}, \hat{H}_{\text{ph}}] = 0.$$

Thus, the interaction preserves the initial state up to a phase. Furthermore, the measured variable, $\hat{N}_e$, commutes with the undisturbed Hamiltonian $\hat{H}_e, \hat{H}_{\text{ph}}$ and, therefore, the scattering operator $\hat{S}$, enables quantum non-demolition (QND) [47–52] measurements of free electron number.

Next, we provide specific examples for the implication of Eq. (5) on states of a populated photonic mode and show the emergence of the known electron-phase retardation in the limit of a classical field.

### III. DISPERSIVE ELECTRON COUNTING

To have a concrete scenario in mind, consider the illustration in Fig. 1(a), where electrons traverse the evanescent field of an optical mode of the microresonator. Assume that any phase-matching terms are nullified by a mismatch between the electron and optical-phase velocities. Thus, the interaction process result from phase-matching independent terms is described by scattering operator in Eq. (5).

Since the scattering of initial state results in a pure phase, the transformation from $|\Psi_i\rangle$ to $|\Psi_f\rangle$ is unobservable without a reference. Mathematically, it can be quantified by the projection of final state on the unperturbed state. Thus, the resulting photonic state, $|\psi_{\text{ph,f}}\rangle$, can be defined as the projection of the combined state on the initial electron state. Considering a state with precisely $N_e$ electrons,

$$|\psi_{\text{ph,f}}\rangle = \langle \psi_e | \hat{S} | \Psi_i \rangle = e^{-i\frac{g_\phi}{\hbar\omega} N_e \hat{H}_{\text{ph}}} |\psi_{\text{ph}}\rangle.$$

By substituting the Schrödinger equation $\hat{H}_{\text{ph}} = i\hbar \partial_t$, This photonic scattering operator can be represented as a time-shifting operator,

$$\hat{S} = e^{\tau \partial_t}.$$

Propagating the photonic states by $\tau = \frac{1}{\omega} g_\phi N_e$. For modes with an angular frequency $\omega$, it is a rotation $\omega \tau = g_\phi N_e$ in phase-space. This optical phase is illustrated in Fig. 1(b).

If the initial photon state is a coherent state $|\alpha\rangle$, for $N_e$ electrons, $|\psi_{\text{ph,f}}\rangle = |\alpha e^{-ig_\phi N_e}\rangle$, the optical state shifts

$$|\alpha\rangle \rightarrow |\alpha e^{-ig_\phi N_e}\rangle. \quad (7)$$

The transformation adds a phase to the coherent-state parameter, which is a pure rotation in phase space, $g_\phi N_e$, as shown in Fig. 1(c). In other words, a classical light field populating a single photonic mode will undergo a phase shift of $-g_\phi$ for every electron passage! For a beam with many electrons, the optical phase shift, $\delta\phi$, accumulates linearly, yielding a proportion to the electron number (detail derivations see Appendix D),

$$\delta\phi = -g_\phi N_e. \quad (8)$$

In principle, this enables electron counting by measuring laser phase-shift, as schematically shown in Fig. 1(b). We show in a later part that this is state preserving, that is, if $N_e$ electrons pass within the mode's coherence time, they maintain their initial state.

One can apply the elastic scattering operator to additional photonic states. If the initial state is squeezed vacuum $|0,\zeta\rangle$ [53], where $\zeta$ is squeezing parameter, $\hat{S}$ induces the transformation (detail derivations see Appendix E):

$$|0,\zeta\rangle \rightarrow |0,\zeta e^{-2ig_\phi N_e}\rangle. \quad (9)$$

The factor $e^{-2ig_\phi N_e}$ indicates that the free electrons can rotate the squeezing direction via elastic interaction, as shown in Fig. 1(d). For large number electrons, it could be possible to rotate the squeezing direction with $\pi/2$.

For coherent squeezed state $|\alpha,\zeta\rangle$ [54,55], $\hat{S}$ induces the transformation:

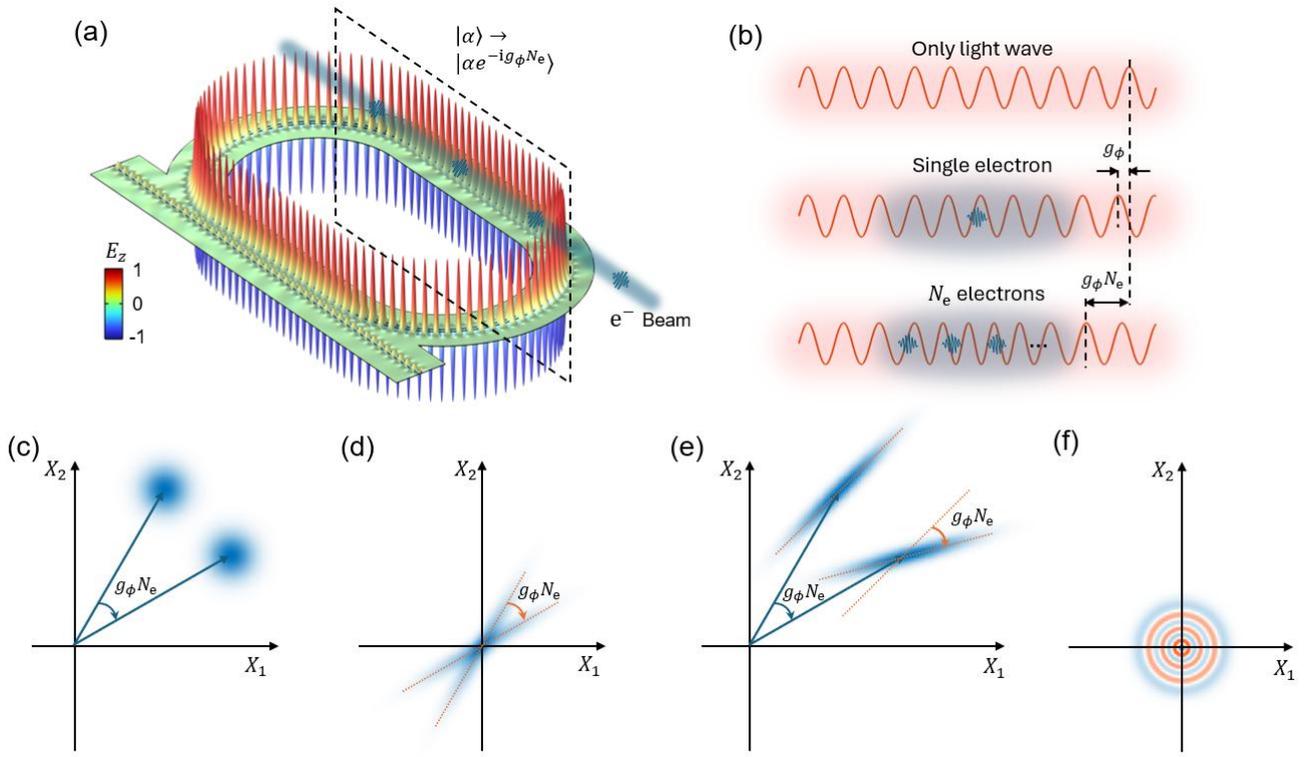

FIG.1. Optical phase shifts induced by elastically scattered free electrons. (a) A schematic shows the interaction of free electrons with a phase-mismatching laser on racetrack microresonator. The phase shift added to a coherent state $|\alpha\rangle$ is noted. (b) The optical phase shift is proportional to the electron number, $N_e$. The red sinusoidal curves represent light waves. (c-e) Phase shifts of Gaussian states represented in phase space by their Wigner functions: (c) coherent state for $|\alpha| = 5$, (d) squeezed vacuum state ($|\zeta| = 1$), (e) coherent squeezed state ($|\alpha| = 5, |\zeta| = 1$): (f) non-Gaussian number state ($N = 6$).

$$|\alpha, \zeta\rangle \rightarrow |\alpha e^{-ig_\phi N_e}, \zeta e^{-2ig_\phi N_e}\rangle.$$

This is a combined results of Eq. (7) and Eq. (9). Hence, the elastic interaction of coherent squeezed state with free electrons adds a phase shift of the electric field and a rotation of squeezing parameter, as shown in Fig. 1(e).

If the initial state is the joint state of two modes of coherent states, such as cat state [56–58], assuming the elastic coupling strength of these two modes equal,

$$|\alpha\rangle + e^{i\theta}|-\alpha\rangle \rightarrow |\alpha e^{-ig_\phi N_e}\rangle + e^{i\theta}|-\alpha e^{-ig_\phi N_e}\rangle,$$

which is the abovementioned rotation in phase space around the origin for both modes.

If the initial photon state is a Fock state, $|N\rangle$, the operator $\hat{S}$ just adds a phase, i.e.,

$$|N\rangle \rightarrow e^{-ig_\phi N_e N}|N\rangle,$$

equivalent to $\omega\tau$. In this trivial scenario, the interaction with the electron adds a pure global phase since the number state is a stationary state (Fig. 1(f)).

One can consider N00N state, namely, a Fock states with N photons in one or the other arm of an interferometer [5,59], where the free electron interacts in one of the arms only. The transformation is then

$$\tfrac{1}{\sqrt{2}}(|N\rangle|0\rangle + |0\rangle|N\rangle) \rightarrow \tfrac{1}{\sqrt{2}}\big(e^{-ig_\phi N_e N}|N\rangle|0\rangle + |0\rangle|N\rangle\big).$$

A subsequent interference between the two arms will oscillate proportionally to the multiplication $N_e N$.

## IV. THE ELECTRON-PHASE RETARDER AS THE CLASSICAL LIMIT

In a similar treatment, one can address the effect of the elastic scattering operator in Eq. (5) on the electrons, for a specific initial photonic state, $|\psi_{\text{ph}}\rangle$. The final electron state is then $|\psi_{e,f}\rangle = \langle\psi_{\text{ph}}|\hat{S}|\Psi_i\rangle$. For an initial Fock state, $|\psi_{\text{ph}}\rangle = |N\rangle$, the final electronic state is

$$|\psi_{e,f}\rangle = \prod_k e^{-ig_\phi N} \hat{c}_k^\dagger |0\rangle_e.$$

Thus, electrons phase shift is proportional to photon numbers.

To enable the investigation of classical fields, we choose $|\psi_{\text{ph}}\rangle = |\alpha\rangle$. The scattering operator $\hat{S}$ gives system final state $|\Psi_f\rangle = |\alpha e^{-ig_\phi N_e}\rangle \otimes |\psi_e\rangle$. The projection on the initial optical state in the case of an optical coherent state is $\langle\alpha|\alpha e^{-ig_\phi}\rangle =$

$\exp\left(|\alpha|^2(e^{-ig_\phi}-1)\right) \approx e^{-ig_\phi|\alpha|^2}$ for small $g_\phi$. Thus, we find

$$|\psi_{e,f}\rangle = \prod_k e^{-ig_\phi|\alpha|^2} \hat{c}_k^\dagger|0\rangle_e.$$

Since $|\alpha|^2 = \bar{N}$ is the average photon number in coherent state, the electron phase shifts by $-g_\phi \bar{N}$. At the limit of a classical field, where $\bar{N} \gg 1$, our quantum framework reproduces the known effect, where a strong laser field acts as a local phase retarder for free electrons, resulting in the Kapitza-Dirac effect [35–37] and laser-based Zernike holography [38–41]. It should be noted that the global phase we omitted from Eq. (5), $e^{i(\chi-\frac{1}{2}g_\phi)}$, could be relevant in comparison to the phase $g_\phi \bar{N}$, that results from weak laser fields.

## V. THE REFRACTIVE INDEX OF FREE ELECTRONS

To write the effective refractive index of the electron, we first rewritten the elastic coupling amplitude defined in Eq. (4) with $\mathbf{\Gamma} = \text{diag}\left(1,1,\frac{1}{\gamma^2}\right)$. Assuming electrons are propagating along z-axis, and the transverse position, $\mathbf{r}_\perp = (x,y)$, $g_\phi$ can be expressed as a function of the optical field integrated over the electron trajectory,

$$g_\phi(\mathbf{r}_\perp) = \frac{e^2}{\hbar p_0 \omega^2}\int_0^L dz\left(|\mathbf{E}_\perp(r)|^2 + \frac{1}{\gamma^2}|E_z(r)|^2\right). \quad (10)$$

To estimate its figure of merit, we substitute the classical radius of the electron to the above equation, $r_e = \frac{e^2}{4\pi\epsilon_0 m_e c^2} \approx 2.8$ fm, such that its pre-factor can be rewritten as $\frac{e^2}{p_0\hbar\omega^2} = \frac{2r_e}{\gamma\beta}\frac{\lambda_0\epsilon_0}{\hbar\omega}$. Here, $\epsilon_0$ is the vacuum permittivity and $\lambda_0 = \frac{2\pi c}{\omega}$ is the optical wavelength in vacuum. Assuming a mode volume $V$ and, for now, polarization perpendicular to electron propagation direction, the photonic electric field is $|\mathbf{E}| = \sqrt{\frac{\hbar\omega}{\epsilon_0 V}}$. Thus, the elastic coupling strength reduces to a comparison between natural scales of the electron and optical quanta,

$$g_\phi = \frac{2r_e/(\gamma\beta)}{V/(\lambda_0 L)} = \frac{\text{electron diameter}}{\text{photon diameter}}. \quad (11)$$

The electron diameter, $\frac{2r_e}{\gamma\beta}$, is its classical diameter subjected to the relativistic contraction factor. The photon diameter, $\frac{V}{\lambda_0 L}$, is defined by photon mode volume divided by the wavelength and interaction length. Since the classical radius of the electron is much smaller than the optical wavelength, this natural discrepancy of scales determines the weakness of the elastic interaction between an electron and a photon. For a photon scale of a wavelength with mode volume $V = \lambda_0^3$, interaction length $L = \lambda_0$, and $\lambda_0 = 1550$nm, $g_\phi$ is in an order of $10^{-7}$ to $10^{-10}$. Fig. 2 shows the dependence of $g_\phi$ on kinetic energy of the electron.

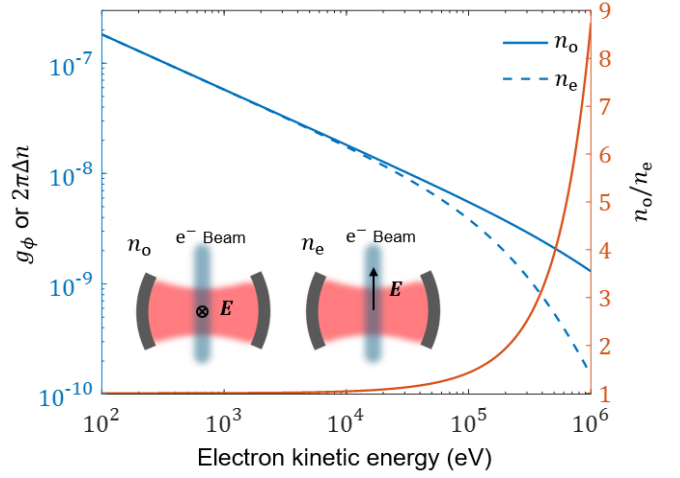

FIG.2. Elastic electron-photon coupling amplitude (i.e., refractive index) versus electron energy. The birefringence of free electrons as optical material becomes apparent for kinetic energies above $10^5$ eV. The insert illustrates the ordinary and extraordinary polarization of the electron birefringence. The numerical values here assume the light is confined in an optical cavity with a volume $V = \lambda_0^3$, interaction length $L = \lambda_0$ and $\lambda_0 = 1550$nm.

The refractive index of the electron can be inferred from the optical phase it induced in Eq. (8), $\delta\phi$, and the length of its trajectory, by $\delta\phi = -\frac{2\pi}{\lambda_0}\int dz\,\delta n$. Although arbitrary, one can use coherent states and Eq. (8) to write $\delta n = N_e \frac{\lambda_0}{2\pi}\frac{d}{dz}g_\phi$. Furthermore, combining Eq. (10), we have

$$\delta n = N_e \frac{e^2 c}{p_0 \hbar \omega} \mathbf{A}^\dagger \mathbf{\Gamma} \mathbf{A} = N_e \frac{e^2 c}{p_0 \hbar \omega^3}\left(|\mathbf{E}_\perp|^2 + \frac{1}{\gamma^2}|E_z|^2\right). \quad (12)$$

Thus, the electron-induced $\delta n$ is birefringent, governed by the birefringence matrix, $\mathbf{\Gamma}$, and the overlap of the electron with the mode, $\mathbf{E}_\perp, E_z$. The optical axis is along the electron velocity. The ratio of ordinary and extraordinary refractive indices is $\gamma^{-2}$, as shown in Fig. 2, becomes significant for kinetic energy scales of transmission electron microscopes, above 100 keV. The electron refractive index is proportional to $N_e$, the number of electrons that arrive within the lifetime of the optical mode, in agreement with our previous referral to the dispersive photonic Hamiltonian.

As a final note, in the simplified conceptual scenario where all the photonic dimensions are $\lambda_0$, with uniformly distributed light fields, Eq. (12) reduces to $\delta n = \frac{1}{2\pi} N_e g_\phi$, with $g_\phi$ as in Eq. (11). In other words, the figure of merit for the electron's refractive index is

$$\delta n = \frac{N_e}{2\pi} \frac{\text{electron diameter}}{\text{photon diameter}}.$$

## VI. PROPOSED EXPERIMENTAL REALIZATION

We address the challenge of counting free electrons by measuring extremely small phase shifts using interferometry with an optical cavity. The motivation is the prospect of superb spectral precision in microresonators, that may enable a sensitivity necessary for the detection of a single electron. Furthermore, high quality optical resonators can be fabricated with various dielectric materials and geometrical shapes [60–62].

Figure 3a outlines the schematics of QND free-electron counting in electron microscope. A racetrack resonator is placed in the vacuum chamber and connected with input/output fibers. If a circulating optical mode is strongly confined, an electron-induced intra-resonator phase shift changes the output power. For a high finesse ($\mathcal{F}$) racetrack resonator, which is available in modern micro/ nanofabrication technologies [63–65], the power in the racetrack ($P_{\text{cir}}$) can be enhanced by several orders of magnitude compared with the input power ($P_{\text{inc}}$). The enhancement factor (see Fig. 3b) depends on phase accumulated in one round-trip [66], $\phi_R = T_R \omega$, where $T_R$ is the duration of one round trip. An intra-cavity phase added dynamically by a traversing electron is equivalent to a time-varying detuning. The sensitivity is optimal when the cavity is pumped on the side of the resonance peak, where the derivative of the enhancement factor is maximal.

At this spectral point, near the half-maximum points of the resonance peak, the enhancement factor is approximately $\frac{\mathcal{F}^2}{\pi^2}$ (see Appendix F). Such side-of-fringe locking method were popular for laser frequency stabilization before the interferometric PDH approach by Pound Drever and Hall [67–69].

Since the effect induced by the electron is transient, we turn to a dynamical analysis of the optical effects induced in the cavity. The energy variation in the cavity depends on two of its timescales, the roundtrip ($T_R$) and the ringdown time (or lifetime). We refer to electron's passage as time-zero and analyze the light in the cavity in subsequent roundtrips.

The full expressions of circulating field and output field with respect to their steady state are derived in Appendix F. To outline the dynamics, consider the optical amplitude and phase accumulated in a round trip, expressed as $x = ra e^{i\phi_R}$, $r$ is the coefficient of light to remain in resonator and $a$ is the transmission, such that $(1 - a^2)$ is the absorption in one roundtrip [66]. An abrupt phase-shift in the resonator generates a pulse train in the circulating and output fields ringing with intervals $T_R$, where the $m^{\text{th}}$ ringing event has a complex amplitude $x^m$, decaying per roundtrip. The effects of phase shift will be added in every one-round trip and the changes of circulating and output light fields is proportional to the sum of every round trip, that is $\Sigma_{m=-\infty}^{+\infty} x^{m+1}(e^{i\delta\phi(t-mT_R)} - 1)$, where $m$ is an integer. For resonators with a high-quality factor, the decay is slow, and the output would have many weak echoes of the phase-shift pulse (see Fig. 5).

For simplicity, we consider the scenario of phase shift duration larger than $T_R$. The time-dependence of the induced dynamic processes on the circulating power and output power are shown in Fig. 3c. The circulating power decreases at first, then recovers to a steady state due to the refilling of input power. The output power increases at the beginning, within the round-trip time, then decreases to lower than steady state level, until the resonator stabilizes at the scale of the cavity lifetime. The net energy output due to the phase shift is proportional to the first order derivative of the circulating power and the energy absorption in the resonator, $E_{\text{net}} \propto (1 - a^2) \frac{d}{d\omega} P_{\text{cir}}(\omega)$. In a lossless cavity ($a = 1$) the effect nullifies.

To calculate a concrete example and evaluate the electron-counting sensitivity, we assume the electron beam induces a constant phase shift for a duration of $T_{\text{int}}$. The energy output in a high-finesse microresonator is (the detailed derivation is in Appendix F)

$$E_{\text{net}} = -\frac{1}{2} P_{\text{cir}} \cdot T_{\text{int}} \cdot \delta\phi \qquad (13)$$

Therefore, the net energy output for a single electron passage is proportional to the circulating power. For a microresonator with high finesse, one can benefit from the low input power and more importantly, the low background of a weak output power. If we consider a racetrack microresonator with 100-μm-long straight segment, a circulating power of 100 mW, an electron with a kinetic energy of 100 eV, then interaction time is $T_{\text{int}} =$ 17 ps and $g_\phi \approx 1.8 \times 10^{-7}$. The net energy output from resonator according to Eq. (13) is approximately $1.5 \times 10^{-19}$ joule. This value is approximately the energy of two photons per passing electron. Given that this photon may be overwhelmed by noise in the system, it would be challenging to probe the electron-number with a single-electron precision.

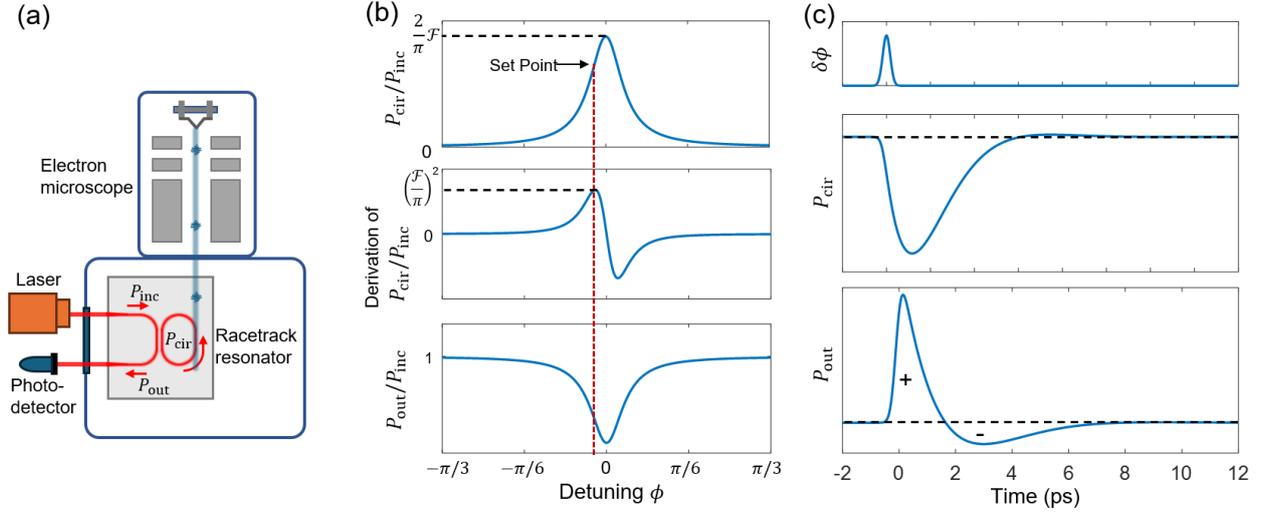

FIG.3. A proposal for measuring electron-induced phase shifts in optical microresonator. (a) Schematic outline of a setup for counting free electrons by measuring the energy output of a cavity. The optical phase shift of light circulating in the resonator due to electron passage leads to the energy leakage, which could be read out. (b) Side-of-fringe locking approach. For high finesse resonators, the intracavity buildup factor (upper panel) is very sensitive to the detuning. Locking the detuning phase at maximum sensitive point (middle panel), the output power also maximally sensitive (bottom panel). (c) Dynamic of the circulating and output power following an electron passage assuming the cavity's round-trip and lifetimes are 0.2 ps and 5 ps, respectively. The upper panel shows a time-dependent Gaussian-shape phase shift caused by one electron, assuming a FWHM transition time of 0.4 ps. It results in a temporary decrease of the circulating power (middle panel) and an increase of the output power (bottom panel). The net energy output is the region marked '+' minus the region marked '-'.

## VII. SUMMARY, DISCUTION AND OUTLOOK

In conclusion, we show that the elastic scattering of electrons and photons is governed by a dispersive Hamiltonian, which has the form of a quantum non-demolition electron counting operator. We provide the theoretical framework for the phase induced by the elastic interaction of electrons and photons, show that it can be quantified as an effective refractive index of an electron, and offer concrete numerical examples. The effect of the electron-induced phase is shown for various photonic states, as coherent states, squeezed states, Fock states, and N00N states. Furthermore, the induced phase is slightly birefringent along the axis of electron propagation. In principle, the linear scaling of the induced optical phase with the electron number could allow the counting of electrons by monitoring the phase shift in a high-finesse optical resonator. This QND measurement may be advantageous since consecutive electrons arriving within the lifetime of the cavity maintain their indistinguishability, thereby providing a good quantum estimator for their number. We provide a concrete example of a high-finesse microresonator that translates the electron-induced phase shift to a net energy output. Although this example is not sufficient to support single-electron sensitivity, its sensitivity could be increased by a high circulating power and long interaction time, or by utilizing a different system altogether

We emphasize that in the dispersive-Hamiltonian regime the phase-shifts of both electrons and photons do not emanate from photon absorption/emission process as in the inelastic interactions, such as photon-induced near-field electron microscopy [16,17] and laser-driven structure-based acceleration [70–73], and hence conserve the initial quantum states. For the same reason, the elastic terms dominate when such energy- and momentum-exchanges are suppressed. The effect's sensitivity for the electron velocity (or acceleration voltage) is related to the time the electron spends in the optical field and relativistic effects reflected in the birefringence matrix. In addition to the fundamental concepts that we present here, if a future work succeeds to push the sensitivity of the dispersive electron-photon scattering towards the single-electron sensitivity, it may open the path to enhanced quantum measurements with electrons, based on quantum electron-number states, alongside the atomic resolution of such microscopes. Its application within electron microscopy may bring about a superb signal-to-noise ratio [74], beyond standard quantum limit.

## ACKNOWLEDGMENTS

This research was supported by The Israel Science Foundation (grant No. 2992/24 and 1021/22) and the Young Faculty Award from the National Quantum Science and Technology program of the Israeli Planning and Budgeting Committee. D. Z. gratefully acknowledges the Heynman Foundation for Post-doc fellowship.

## APPENDIX A: THE HAMILTONIAN OF FREE-ELECTRON-PHOTON SYSTEM

The free-electron-photon interaction Schrödinger equation in different forms has been solved in literatures. For the first order interaction, i.e., interaction Hamiltonian proportional to vector potential $\boldsymbol{A}$, the solutions has been demonstrated in the many works about quantum photon-induced near-field electron microscopy [7,17,27,75]. For the second order interaction, i.e., interaction Hamiltonian proportional to $\boldsymbol{A}^2$, also has been solved in the studies of laser manipulation of electron beam [42,45,43,76]. However, the full quantum theory that involves both free electron system and photon system and their interaction to second order is lacking.

To show the effect of free-electron-photon interaction on photon system, let start with the consideration of a free electron with momentum $\boldsymbol{p}_0$ and rest energy $m_{\mathrm{e}}c^2$, propagates in free space. A concrete example of such scenario is the flying free electrons as probes in electron microscopes. The relativistic energy of electron is

$$\varepsilon = \sqrt{m_{\mathrm{e}}^2 c^4 + p^2 c^2}. \quad (A1)$$

To expand the electron energy at the initial energy $\varepsilon_0 = \sqrt{m_{\mathrm{e}}^2 c^4 + p_0^2 c^2}$, let's define the changes of momentum $\Delta \boldsymbol{p} = \boldsymbol{p} - \boldsymbol{p}_0$. Replacing momentum in Eq. (A1) by $\boldsymbol{p}_0 + \Delta \boldsymbol{p}$ and substitute $\varepsilon_0$, we have

$$\varepsilon = \varepsilon_0 \sqrt{1 + \tfrac{2}{\varepsilon_0} \boldsymbol{v}_0 \cdot \Delta \boldsymbol{p} + \tfrac{c^2}{\varepsilon_0^2} (\Delta \boldsymbol{p}) \cdot (\Delta \boldsymbol{p})}.$$

where $\boldsymbol{v}_0 = \boldsymbol{p}_0 c^2 / \varepsilon_0$ is electron initial group velocity. Using formular $\sqrt{1+x} = 1 + \tfrac{x}{2} - \tfrac{x^2}{8} + \cdots$ to expand the electron energy as a function of $\Delta \boldsymbol{p}$ to second order term, then the electron energy can be written as

$$\varepsilon(\Delta \boldsymbol{p}) = \varepsilon_0 + \boldsymbol{v}_0 \cdot \Delta \boldsymbol{p} + \frac{1}{2\gamma m_{\mathrm{e}}} (\Delta \boldsymbol{p})^{\mathrm{T}} \boldsymbol{\Gamma} (\Delta \boldsymbol{p}) + \cdots \quad (A2)$$

where $\boldsymbol{\Gamma}$ is a 3×3 matrix and defined as

$$\boldsymbol{\Gamma} = I_3 - \frac{\boldsymbol{v}_0}{c} \frac{\boldsymbol{v}_0^{\mathrm{T}}}{c},$$

$I_3$ is a $3 \times 3$ identity matrix. If electrons propagate along z-direction, then $\boldsymbol{v_0} = \begin{bmatrix} 0 \\ 0 \\ v_0 \end{bmatrix}$, $\boldsymbol{\Gamma}$ reduce to a diagonal matrix and

$\boldsymbol{\Gamma} = \mathrm{diag}(1,1,\tfrac{1}{\gamma^2})$.

In the other hand, consider a monochromatic light propagates in vacuum or an optical microcavity. According to the theory of quantum optics, the vector potential of light can be written as

$$\widehat{\boldsymbol{A}} = \boldsymbol{A} \hat{a} \mathrm{e}^{-\mathrm{i}\omega t} + \boldsymbol{A}^* \hat{a}^{\dagger} \mathrm{e}^{\mathrm{i}\omega t}, \quad (A3)$$

where $\boldsymbol{A} = -\mathrm{i}\tfrac{1}{\omega} \boldsymbol{E}(\boldsymbol{r})$, $\boldsymbol{A}^* = \mathrm{i}\tfrac{1}{\omega} \boldsymbol{E}^*(\boldsymbol{r})$, $\omega$ is light frequency, $\hat{a}$ ($\hat{a}^{\dagger}$) is photon annihilation (creation) operator, $\boldsymbol{E}(\boldsymbol{r})$ is the electric field of single photon. The Hamiltonian of light is

$$\widehat{H}_{\mathrm{ph}} = \hbar \omega \left( \hat{a}^{\dagger} \hat{a} + \frac{1}{2} \right) \quad (A4)$$

In the free-electron-photon interaction, consider electrons and photons as a combined system that the joint state can be written as $|\psi_{\mathrm{ph}}\rangle \otimes |\psi_{\mathrm{e}}\rangle$. Working at minimal coupling scheme, the momentum $\boldsymbol{p}$ can be replaced by mechanical momentum operator $\widehat{\boldsymbol{p}} + e\widehat{\boldsymbol{A}}$, where $\widehat{\boldsymbol{p}} = -\mathrm{i}\hbar \nabla$ is the momentum operator, $e$ is elemental charge. Since it has no charges or current in the path of electron, the scalar potential of electromagnetic wave is zero. Ignoring the spin of electron, taking the expression of $\Delta \boldsymbol{p} = \widehat{\boldsymbol{p}} - \boldsymbol{p}_0 + e\widehat{\boldsymbol{A}}$ into Eq. (A2), we have the Hamiltonian of electron-photon combined system,

$$\widehat{H} \cong \widehat{H}_{\mathrm{ph}} + \varepsilon_0 + \boldsymbol{v}_0 \cdot (\hat{p} - \boldsymbol{p}_0 + e\widehat{\boldsymbol{A}})$$
$$+ \frac{1}{2\gamma m_{\mathrm{e}}} (\widehat{\boldsymbol{p}} - \boldsymbol{p}_0 + e\widehat{\boldsymbol{A}})^{\mathrm{T}} \boldsymbol{\Gamma} (\widehat{\boldsymbol{p}} - \boldsymbol{p}_0 + e\widehat{\boldsymbol{A}}). \quad (A5)$$

Eq. (A5) represents the Hamiltonian of single electron with light field. For counting electrons, the quantization of electrons is necessary. By using the fermion annihilation (creation) operator $\hat{c}_k$ ($\hat{c}_k^{\dagger}$), which satisfied anticommutation relation $\{\hat{c}_{k'}, \hat{c}_k^\dagger\} = \delta_{k',k}$ and $\{\hat{c}_{k'}, \hat{c}_k\} = \{\hat{c}_{k'}^\dagger, \hat{c}_k^\dagger\} = 0$, the Hamiltonian in Eq. (A5) can be modified and applied to multi-electron system. Separating the system's Hamiltonian into non-interaction term and interaction term,

$$\widehat{H} = \widehat{H}_0 + \widehat{H}_{\text{int}}, \tag{A6}$$

where

$$\widehat{H}_0 = \widehat{H}_{\text{ph}} + \widehat{H}_e. \tag{A7}$$

$\widehat{H}_{\text{ph}}$ is defined in Eq. (A4) and the non-interaction Hamiltonian of electrons

$$\widehat{H}_e = \sum_{k=k_1}^{k_{N_e}} \hat{\varepsilon}_k \hat{c}_k^\dagger \hat{c}_k \tag{A8}$$

where $\varepsilon_k$ is single electron Hamiltonian. According to Eq. (A5),

$$\hat{\varepsilon}_k = \varepsilon_{0,k} + \boldsymbol{v}_{0,k} \cdot (\hat{p} - \boldsymbol{p}_{0,k}) + \frac{1}{2\gamma_k m_e}(\hat{\boldsymbol{p}} - \boldsymbol{p}_{0,k})^{\text{T}} \boldsymbol{\Gamma}(\hat{\boldsymbol{p}} - \boldsymbol{p}_{0,k}).$$

Since we care about the phase mismatching interaction, the evanescent variations of electron energy during the interaction would be small. Therefore, the second order term in $\hat{\varepsilon}_k$ presented the square of momentum changes can be neglected. Furthermore, if we consider all electrons in one mode, then $\hat{\varepsilon}_k$ reduce to

$$\hat{\varepsilon} = \varepsilon_0 + \boldsymbol{v}_0 \cdot (\hat{p} - \boldsymbol{p}_0). \tag{A9}$$

The interaction Hamiltonian

$$\widehat{H}_{\text{int}} = \sum_{k=k_1}^{k_{N_e}} \left( e\boldsymbol{v}_0 \cdot \widehat{\boldsymbol{A}} + \frac{e^2}{2\gamma m_e}\widehat{\boldsymbol{A}}^{\text{T}}\boldsymbol{\Gamma}\widehat{\boldsymbol{A}} + \widehat{H}_{pA} \right)\hat{c}_k^\dagger \hat{c}_k \tag{A10}$$

where coupling term $\widehat{H}_{pA}$ is linear with vector potential and dependent on the changes of momentum,

$$\widehat{H}_{pA} = -\frac{e}{2\gamma m_e}\left((\hat{\boldsymbol{p}} - \boldsymbol{p}_0)^{\text{T}}\boldsymbol{\Gamma}\widehat{\boldsymbol{A}} + \widehat{\boldsymbol{A}}^{\text{T}}\boldsymbol{\Gamma}(\hat{\boldsymbol{p}} - \boldsymbol{p}_0)\right).$$

Choose Coulomb gauge ($\nabla \cdot \boldsymbol{A} = 0$), $\widehat{H}_{pA}$ can be reduced to $-\frac{e}{\gamma m_e}\widehat{\boldsymbol{A}}^{\text{T}}\boldsymbol{\Gamma}(\hat{\boldsymbol{p}} - \boldsymbol{p}_0)$. Since the changes of momentum are small at phase-mismatching interaction, we neglect $\widehat{H}_{pA}$ in the following calculation. Take Eq. (A3) into Eq. (A10) and rearrange, we have the interaction Hamiltonian

$$\widehat{H}_{\text{int}} = \sum_{k=k_1}^{k_{N_e}} \left(\widehat{H}_1 + \widehat{H}_2 + \widehat{H}_\phi\right)\hat{c}_k^\dagger \hat{c}_k \tag{A11}$$

where $\widehat{H}_1$, $\widehat{H}_2$ and $\widehat{H}_\phi$ are defined in Eq. (2a), (2b) and (2c).

## APPENDIX B: THE EXPRESSION OF SCATTERING OPERATOR

Since we consider all electrons in one mode and assume electrons in number states, we can remove the fermion annihilation and creation operator in this section and add it in the final solution as shown in Eq. (3).

The initial state of free-electron-photon combined system, $|\Psi_i\rangle = |\psi_e\rangle \otimes |\psi_{\text{ph}}\rangle$, satisfied

$$i\hbar \frac{\partial}{\partial t}|\Psi_i\rangle = \widehat{H}_0|\Psi_i\rangle,$$

where $\widehat{H}_0$ is defined in Eq. (A7). To solve the interaction of free electrons with photons, working in interaction picture is convenient. In interaction picture, the interaction Hamiltonian should have the transformation $e^{\frac{i}{\hbar}\widehat{H}_0(t-t_0)}\widehat{H}_{\text{int}}(t)e^{-\frac{i}{\hbar}\widehat{H}_0(t-t_0)}$. Because the Eq. (A3) already has the form of interaction picture, thus, the transformation only needs consider the unperturbed Hamiltonian of electrons,

$$\widehat{H}_{\text{int}}(\boldsymbol{r},t) \rightarrow e^{\frac{i}{\hbar}\widehat{H}_e(t-t_0)}\widehat{H}_{\text{int}}(\boldsymbol{r},t)e^{-\frac{i}{\hbar}\widehat{H}_e(t-t_0)} \tag{B1}$$

Taking the expressions in Eq. (A8) and (A9) to above transformation, then $\widehat{H}_{\text{int}}$ should be transformed to $e^{\frac{i}{\hbar}\boldsymbol{v}_0\cdot\hat{p}(t-t_0)}\widehat{H}_{\text{int}}(\boldsymbol{r},t)e^{-\frac{i}{\hbar}\boldsymbol{v}_0\cdot\hat{p}(t-t_0)}$. Because the transformation of position operator, $e^{\frac{i}{\hbar}\boldsymbol{v}_0\cdot\hat{p}t}\boldsymbol{r}e^{-\frac{i}{\hbar}\boldsymbol{v}_0\cdot\hat{p}t}$, equal to $\boldsymbol{r} + \boldsymbol{v}_0 t$, then the transformation in Eq. B1 can be converted to

$$\widehat{H}_{\text{int}}(\boldsymbol{r},t) \rightarrow \widehat{H}_{\text{int}}(\boldsymbol{r} + \boldsymbol{v}_0 t, t).$$

Furthermore, since we assume electron propagate along z-axis and only vector potential in interaction Hamiltonian is dependent on $\boldsymbol{r}$, this result is equal to let the vector potential have a transform

$$\boldsymbol{A}(x,y,z,t) \rightarrow \boldsymbol{A}(x,y,z+v_0t,t). \tag{B2}$$

Thus, the final equivalent result is the Lorentz transformation of vector potential that assumes electrons keep initial velocities.

The evolution of system state from $t_0$ to $t$ is governed by time-evolution operator $\widehat{U}$, which can be written as

$$|\Psi(t)\rangle = \widehat{U}(t,t_0)|\Psi(t_0)\rangle.$$

If $t_0$ trend to minus infinity, $|\Psi(t_0 \rightarrow -\infty)\rangle = |\Psi_i\rangle$. The time-evolution operator satisfied

$$i\hbar\frac{\partial}{\partial t}\widehat{U}(t,t_0) = \widehat{H}_{\text{int}}\widehat{U}(t,t_0).$$

To solve this linear derivation equation, we can work in time-ordered expansion. A direct integral gives the time-ordered exponential

$$\widehat{U}(t,t_0) = \mathcal{T}\exp\left(-\frac{i}{\hbar}\int_{t_0}^{t}dt'\,\widehat{H}_{\text{int}}(z'(t'),t')\right).$$

Here, $\mathcal{T}$ is time-ordering operator. According to Eq. (B2), $z'$ as the function of time can be expressed as $z' = z_0 + v_0(t'-t_0)$. The time-integral can be transformed to position-integral by replacing $t'$ with retarded time, i.e., $t' = t_0 - \left(\frac{z_0}{v_0} - \frac{z'}{v_0}\right)$. Then $dt' = \frac{dz'}{v_0}$, the evolution operator is the function of position,

$$\hat{U}(z, z_0) = \mathcal{T} \exp\left(-\frac{i}{\hbar}\int_{z_0}^{z}\frac{dz'}{v_0}\hat{H}_{\text{int}}(z', t'(z'))\right). \quad (B3)$$

Taking the expression of interaction Hamiltonian in Eq. (2a) into Eq. (B3), the integral of first order interaction Hamiltonian $\hat{H}_1$ in the exponential is

$$-\frac{i}{\hbar}\int_{z_0}^{z}\frac{dz'}{v_0}\hat{H}_1(z', t'(z')) = \frac{i}{\hbar}\int_{z_0}^{z} dz'\left(eA_z(z')\hat{a}e^{-i\omega t'} + \text{c.c.}\right).$$

the factor $e^{-i\omega t'} = e^{-i\omega(t-(\frac{z}{v_0}-\frac{z'}{v_0}))} = e^{-i\omega\frac{z'}{v_0}}e^{i\omega(\frac{z}{v_0}-t)}$. Let $\hat{b} = e^{i\omega(t-\frac{z}{v_0})}$ and rearrange the equation, then the integral of $\hat{H}_1$

$$= -\frac{i}{\hbar}\left(\int_{z_0}^{z}dz'\, eA_z(z')e^{-i\omega\frac{z'}{v_0}}\hat{b}^{\dagger}\hat{a} + \text{c.c.}\right).$$
$$= -g_{\text{Qu}}\hat{b}^{\dagger}\hat{a} + g_{\text{Qu}}^{*}\hat{b}\hat{a}^{\dagger}. \quad (B4)$$

where the inelastic quantum coupling constant is defined by

$$g_{\text{Qu}} = \frac{e}{\hbar\omega}\int_{z_0}^{z}dz'\, E_z(z')e^{-i\omega\frac{z'}{v_0}}. \quad (B5)$$

Here, the relation of $\boldsymbol{A}(\boldsymbol{r}) = -i\frac{1}{\omega}\boldsymbol{E}(\boldsymbol{r})$ has been used in the derivation of Eq. (B4).

Similarly, take $\hat{H}_2$ defined in Eq. (2b) into the integral and replace $e^{-i\omega t'} = e^{-i\omega\frac{z'}{v_0}}\hat{b}$, we have

$$-\frac{i}{\hbar}\int_{z_0}^{z}\frac{dz'}{v_0}\hat{H}_2(z', t'(z')) =$$
$$-\frac{i}{\hbar}\int_{z_0}^{z}dz'\left(\frac{e^2}{2\gamma v_0 m_e}\boldsymbol{A}^{\mathrm{T}}(z')\boldsymbol{\Gamma}\boldsymbol{A}(z')\hat{a}^2 e^{-2i\omega t'} + \text{c.c.}\right)$$
$$= \frac{1}{2}g_2^{*}\hat{b}^{\dagger 2}\hat{a}^2 - \frac{1}{2}g_2\hat{b}^2\hat{a}^{\dagger 2} \quad (B6)$$

where the second order of coupling constant is defined by

$$g_2^{*} = -\frac{i}{\hbar}\frac{e^2}{p_0}\int_{z_0}^{z}dz'\,\boldsymbol{A}^{\mathrm{T}}(z')\boldsymbol{\Gamma}\boldsymbol{A}(z')e^{-2i\omega\frac{z'}{v_0}}. \quad (B7)$$

The integral of $\hat{H}_\phi$ in the exponential is

$$-\frac{i}{\hbar}\int_{z_0}^{z}\frac{dz'}{v_0}\hat{H}_\phi(z', t'(z'))$$
$$= -\frac{i}{\hbar}\int_{z_0}^{z}\frac{dz'}{v_0}\frac{e^2}{2\gamma m_e}\boldsymbol{A}^{\dagger}\boldsymbol{\Gamma}\boldsymbol{A}(2\hat{a}^{\dagger}\hat{a}+1)$$
$$= -ig_\phi\left(\hat{a}^{\dagger}\hat{a}+\frac{1}{2}\right). \quad (B8)$$

where the phase term is defined by

$$g_\phi = \frac{e^2}{\hbar p_0}\int_{z_0}^{z}dz'\,\boldsymbol{A}^{\dagger}(z')\boldsymbol{\Gamma}\boldsymbol{A}(z'). \quad (B9)$$

Taking the integral of the Hamiltonian interaction term in Eq. (B4), (B6) and (B8), into Eq. (B3), the expression of evolution operator is

$$\hat{U}(z, z_0) = \mathcal{T}\exp\left(g_{\text{Qu}}^{*}\hat{b}\hat{a}^{\dagger} - g_{\text{Qu}}\hat{b}^{\dagger}\hat{a} + \frac{1}{2}g_2^{*}\hat{b}^{\dagger 2}\hat{a}^2 \right.$$
$$\left. -\frac{1}{2}g_2\hat{b}^2\hat{a}^{\dagger 2} - ig_\phi\left(\hat{a}^{\dagger}\hat{a}+\frac{1}{2}\right)\right) \quad (B10)$$

Extend the integral to infinity, we have the scattering operator $\hat{S} = \hat{U}(z \to \infty, z_0 \to -\infty)$.

Furthermore, according to Zassenhaus formula [77], $e^{t(X+Y)} = e^{tX}e^{tY}e^{-\frac{t^2}{2}[X,Y]}e^{\frac{t^3}{6}(2[Y,[X,Y]]+[X,[X,Y]])}\cdots$, scattering operator can be separated to a product of infinity factors and each factors only include $g_{\text{Qu}}$, $g_2$ or $g_\phi$. In the first order approximation, the scattering operator only includes the displacement operator $\widehat{D}(g_{\text{Qu}}\hat{b}) = e^{g_{\text{Qu}}^{*}\hat{b}\hat{a}^{\dagger} - g_{\text{Qu}}\hat{b}^{\dagger}\hat{a}}$, the squeeze operator $\hat{S}(g_2\hat{b}^2) = e^{\frac{1}{2}g_2^{*}\hat{b}^{\dagger 2}\hat{a}^2 - \frac{1}{2}g_2\hat{b}^2\hat{a}^{\dagger 2}}$ and the number operator $\widehat{N} = \hat{a}^{\dagger}a$, that is

$$\hat{S} \cong \mathcal{T}\widehat{D}(g_1\hat{b})\hat{S}(g_2\hat{b}^2)e^{-ig_\phi(\widehat{N}+\frac{1}{2})} \quad (B11)$$

The higher order terms include the product of $g_{\text{Qu}}g_2$, $g_{\text{Qu}}g_\phi$, $g_2g_\phi$ and its complex conjugate. Generally, $g_{\text{Qu}} < 1$, $g_2$ and $g_\phi \ll 1$, thus, we can safely neglect the higher order terms.

## APPENDIX C: GLOBAL PHASE

Using Magnus expansion [78], the general solution in Eq. (14) can be expressed without time-ordering operator,
$$\hat{U}(t, t_0) = \exp(\sum_{l=1}^{\infty}\Omega_l).$$
The $\Omega_l$ is the integral of the interaction Hamiltonian or its commutator,
$$\Omega_1 = -\frac{i}{\hbar}\int_{t_0}^{t}dt_1\,\hat{H}_{\text{int}}(t_1),$$
$$\Omega_2 = \frac{1}{2}\left(-\frac{i}{\hbar}\right)^2\int_{t_0}^{t}dt_1\int_{t_0}^{t_1}dt_2\,[\hat{H}_{\text{int}}(t_1), \hat{H}_{\text{int}}(t_2)],$$
...

If only consider first order term, the result is shown in Eq. (B3), but without the time-ordering operator. Considering the second order term, the commutator of interaction Hamiltonians at different times is

$$[\hat{H}_{\text{int}}(t_1), \hat{H}_{\text{int}}(t_2)]$$
$$= [\hat{H}_1(t_1), \hat{H}_1(t_2)] + [\hat{H}_1(t_1), \hat{H}_2(t_2)] + [\hat{H}_2(t_1), \hat{H}_1(t_2)]$$
$$+ \cdots$$
$$= \underbrace{2\text{i} \,\text{Im}\{(q\boldsymbol{v_0} \cdot \boldsymbol{A}(t_1)\text{e}^{-\text{i}\omega t_1})(q\boldsymbol{v_0} \cdot \boldsymbol{A}^*(t_2))\text{e}^{\text{i}\omega t_2}\}}_{c-\text{number}} + (\ldots)\hat{a}$$
$$+ (\ldots)\hat{a}^\dagger + \cdots$$

Neglect all high order terms and only consider the commutator of first order interaction Hamiltonian, $[\hat{H}_1(t_1), \hat{H}_1(t_2)]$, then we have

$$\Omega_2(t, t_0) = -\text{i}\frac{1}{\hbar^2}\int_{t_0}^t dt_1 \int_{t_0}^{t_1} dt_2\, \text{Im}\{(q\boldsymbol{v_0} \cdot \boldsymbol{A}(t_1)\text{e}^{-\text{i}\omega t_1})(q\boldsymbol{v_0} \cdot \boldsymbol{A}^*(t_2))\text{e}^{\text{i}\omega t_2}\}.$$

Transfer the temporal integral to spatial integral,
$$\Omega_2(z, z_0) =$$
$$-\text{i}\left(\frac{q}{\hbar\omega}\right)^2 \int_{z_0}^z dz_1 \int_{z_0}^{z_1} dz_2\, \text{Im}\left\{E_z(z_1)\text{e}^{-\text{i}\omega\frac{z_1}{v_0}} E_z^*(z_2)\text{e}^{\text{i}\omega\frac{z_2}{v_0}}\right\}.$$

Let's define $\Omega_2 = \text{i}\chi_\omega$, then $\chi_\omega$ is a pure phase term. Extend the integral to infinity, the expression of $\chi_\omega$ is

$$\chi_\omega = -\text{Im}\left\{\int_{-\infty}^{+\infty} dz\, g_{\text{Qu}}^*(z) \frac{d}{dz} g_{\text{Qu}}(z)\right\}. \quad (C1)$$

Here, Eq. (B5) has been used. This result indicates that $\chi_\omega$ is the integral over a closed loop in the complex $g_{\text{Qu}}$-plane. It could be rewritten as a compact form: $\chi_\omega = -\text{Im}\{\oint g_{\text{Qu}}^* dg_{\text{Qu}}\}$. The expression in Eq. C1 demonstrates that $\chi_\omega$ is geometric phase [79].

Photon annihilation or creation operator does not appear in Eq. C1, thus, $\chi_\omega$ is independent on light intensity. Origin from vacuum fluctuation, all the optical modes should be taken into consideration, then the global phase is

$$\chi = \int_0^{+\infty} d\omega\, \chi_\omega.$$

Finally, the scattering operator can be written as

$$\hat{S} = \exp(\text{i}\chi)\exp\left(g_{\text{Qu}}^* \hat{b}\hat{a}^\dagger - g_{\text{Qu}} \hat{b}^\dagger \hat{a} + \frac{1}{2}g_2^* \hat{b}^{\dagger 2}\hat{a}^2 - \frac{1}{2}g_2 \hat{b}^2 \hat{a}^{\dagger 2} - \text{i}g_\phi\left(\hat{a}^\dagger\hat{a} + \frac{1}{2}\right)\right).$$

This scattering operator is only applicable to single electron state. Extending the expression to multi electron state, the expression is shown in Eq. (3).

## APPENDIX D: THE PHASE SHIFT OF COHERENT STATE

If the initial photonic state is coherent state, by using $\text{e}^{-\text{i}g_\phi \hat{N}}|\alpha\rangle = |\alpha \text{e}^{-\text{i}g_\phi}\rangle$, we have the final state

$$|\Psi_f\rangle = \hat{S}|\alpha\rangle \otimes |\psi_e\rangle = \text{e}^{\text{i}(\chi - \frac{1}{2}g_\phi)N_e}|\alpha \text{e}^{-\text{i}g_\phi N_e}\rangle \otimes |\psi_e\rangle.$$

then the final photonic state is

$$\langle\psi_e|\Psi_f\rangle = \text{e}^{\text{i}(\chi - \frac{1}{2}g_\phi)N_e}|\alpha \text{e}^{-\text{i}g_\phi N_e}\rangle.$$

To confirm the phase shift of coherent state light is $-g_\phi N_e$, the electric field is calculated as the following. Taking the usual single-mode electric-field operator, $\hat{E}(t) = \boldsymbol{E}^*(\boldsymbol{r})\hat{a}^\dagger \text{e}^{\text{i}\omega t} + \boldsymbol{E}(\boldsymbol{r})\hat{a}\text{e}^{-\text{i}\omega t}$ into the calculation, the expectation value of electrical field in this phase-shift coherent state is $\langle\hat{E}(t)\rangle =$

$\langle\alpha\text{e}^{-\text{i}g_\phi N_e}|\hat{E}(t)|\alpha\text{e}^{-\text{i}g_\phi N_e}\rangle$. Here, the global factor $\text{e}^{\text{i}(\chi-\frac{1}{2}g_\phi)N_e}$

have been dropping out of expectation values. For the expectation value of annihilation and creation operator, we have $\langle\alpha\text{e}^{-\text{i}g_\phi N_e}|\hat{a}|\alpha\text{e}^{-\text{i}g_\phi N_e}\rangle = \alpha\text{e}^{-\text{i}g_\phi N_e - \text{i}\omega t}$, the electric field expectation becomes

$$\langle\hat{E}(t)\rangle = \boldsymbol{E}^*(\boldsymbol{r})\alpha^*\text{e}^{\text{i}g_\phi N_e + \text{i}\omega t} + \boldsymbol{E}(\boldsymbol{r})\alpha\text{e}^{-\text{i}g_\phi N_e - \text{i}\omega t} = 2\text{Re}(\boldsymbol{E}(\boldsymbol{r})\alpha\text{e}^{-\text{i}\omega t - \text{i}g_\phi N_e}).$$

Thus, it can be confirmed that the additional phase in the electric field is $\delta\phi = -g_\phi N_e$.

## APPENDIX E: THE PHASE SHIFT OF COHERENT SQUEEZED STATE

For coherent squeezed state, to prove $\text{e}^{-\text{i}g_\phi \hat{N}}|\alpha, \zeta\rangle = |\alpha\text{e}^{-\text{i}g_\phi}, \zeta\text{e}^{-2\text{i}g_\phi}\rangle$, recall that a coherent squeezed state can be written as: $|\alpha, \zeta\rangle = \hat{D}(\alpha)\hat{S}(\zeta)|0\rangle$, where $\hat{D}(\alpha) = \text{e}^{\alpha\hat{a}^\dagger - \alpha^*\hat{a}}$ is the displacement operator and $\hat{S}(\zeta) = \text{e}^{\frac{1}{2}(\zeta^*\hat{a}^2 - \zeta\hat{a}^{\dagger 2})}$ is the squeezing operator.

Let us first analyze the transformation of the annihilation operator $\hat{a}$ under conjugation by the phase shift operator $\text{e}^{-\text{i}g_\phi \hat{N}}$. Using the Baker-Campbell-Harsdorf formular: $\text{e}^{\hat{A}}\hat{B}\text{e}^{-\hat{A}} = \hat{B} + [\hat{A}, \hat{B}] + \frac{1}{2!}[\hat{A}, [\hat{A}, \hat{B}]] + \cdots$, we have

$$\text{e}^{-\text{i}g_\phi \hat{N}}\hat{a}\text{e}^{\text{i}g_\phi \hat{N}} = \hat{a} + \text{i}g_\phi \hat{a} + \frac{1}{2!}(\text{i}g_\phi)^2 \hat{a} + \cdots = \text{e}^{\text{i}g_\phi}\hat{a}.$$

Similarly, we obtain:
$$\text{e}^{-\text{i}g_\phi \hat{N}}\hat{a}^2 \text{e}^{\text{i}g_\phi \hat{N}} = \text{e}^{-\text{i}g_\phi \hat{N}}\hat{a}\text{e}^{\text{i}g_\phi \hat{N}}\text{e}^{-\text{i}g_\phi \hat{N}}\hat{a}\text{e}^{\text{i}g_\phi \hat{N}}$$
$$= (\text{e}^{\text{i}g_\phi}\hat{a})^2 = \text{e}^{2\text{i}g_\phi}\hat{a}^2.$$
$$\text{e}^{-\text{i}g_\phi \hat{N}}\hat{a}^\dagger \text{e}^{\text{i}g_\phi \hat{N}} = \text{e}^{-\text{i}g_\phi}\hat{a}^\dagger,$$
$$\text{e}^{-\text{i}g_\phi \hat{N}}\hat{a}^{\dagger 2}\text{e}^{\text{i}g_\phi \hat{N}} = \text{e}^{-2\text{i}g_\phi}\hat{a}^{\dagger 2}.$$

Now consider the transformation of the displacement operator:
$$\text{e}^{-\text{i}g_\phi \hat{N}}\hat{D}(\alpha)\text{e}^{\text{i}g_\phi \hat{N}} = \exp(\alpha\text{e}^{-\text{i}g_\phi}\hat{a}^\dagger - \alpha^*\text{e}^{\text{i}g_\phi}\hat{a}) = \hat{D}(\alpha\text{e}^{-\text{i}g_\phi}).$$

The transformation of the squeezing operator:
$$\text{e}^{-\text{i}g_\phi \hat{N}}\hat{S}(\zeta)\text{e}^{\text{i}g_\phi \hat{N}} = \exp\frac{1}{2}(\zeta^*\text{e}^{2\text{i}g_\phi}\hat{a}^2 - \zeta\text{e}^{-2\text{i}g_\phi}\hat{a}^{\dagger 2}) =$$

$\hat{S}(\zeta e^{-2ig\phi}).$

Finally, applying the transformed operators to the coherent squeezed state:

$$e^{-ig\phi\hat{N}}|\alpha,\zeta\rangle = e^{-ig\phi\hat{N}}\hat{D}(\alpha)\hat{S}(\zeta)|0\rangle =$$
$$e^{-ig\phi\hat{N}}\hat{D}(\alpha)e^{ig\phi\hat{N}}e^{-ig\phi\hat{N}}\hat{S}(\zeta)e^{ig\phi\hat{N}}e^{-ig\phi\hat{N}}|0\rangle =$$
$$\hat{D}(\alpha e^{-ig\phi})\hat{S}(\zeta e^{-2ig\phi})|0\rangle = |\alpha e^{-ig\phi},\zeta e^{-2ig\phi}\rangle.$$

Thus, we have the transformation of coherent squeezed state. Let $\alpha = 0$, then have the transformation of squeezed vacuum state.

## APPENDIX F: STATIC ANALYSIS OF OPTICAL MICRORESONATOR

In a single optical all-pass microresonator, the basic relation among the incident field $E_1 e^{i\omega t}$ (the incident power $P_{\text{inc}} \propto |E_1 e^{i\omega t}|^2$), the output field $E_2 e^{i\omega t}$ (the output power $P_{\text{out}} \propto |E_2 e^{i\omega t}|^2$), the circulating field just before the coupling point $E_3 e^{i\omega t}$ (the circulating power $P_{\text{cir}} \propto |E_3 e^{i\omega t}|^2$) and after the coupling point $E_4 e^{i\omega t}$ are related via following unitary matrix [66]:

$$\begin{pmatrix} E_2 \\ E_4 \end{pmatrix} = \begin{pmatrix} r & i\kappa \\ i\kappa & r \end{pmatrix} \begin{pmatrix} E_1 \\ E_3 \end{pmatrix} \quad (F1)$$

where the lumped self- and cross-coupling coefficients $r$ and $\kappa$ satisfy the relation $r^2 + \kappa^2 = 1$, both $r$ and $\kappa$ are real number. At the steady state, the circulating fields $E_3$ and $E_4$ are connected by round-trip relation: $E_3 = a e^{i\phi_R} E_4$. Here, $\phi_R = \omega T_R$, $T_R$ is the round-trip time of light in resonator. $a$ represents the single-pass amplitude transmission and $0 < a < 1$. Solve the matrix in Eq. (F1), we have the relation at the steady state:

$$E_{4,s} = \frac{i\kappa}{1-x} E_1, \quad (F2)$$

$$E_{2,s} = \frac{r - a e^{i\phi_R}}{1-x} E_1, \quad (F3)$$

$$E_{3,s} = a e^{i\phi_R} E_{4,s}. \quad (F4)$$

where $x = r a e^{i\phi_R}$ is the one round trip factor.

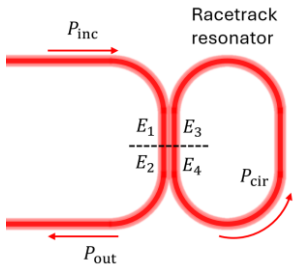

FIG. 4. Schematically show the incident field ($E_1$), the output field ($E_2$) and the circulating fields ($E_3 \& E_4$).

The ratio of circulating power to incident power, or the buildup factor $B$, can be expressed as the function of normalized detuning

$$B(\phi_R) = \frac{P_{\text{cir}}(\phi_R)}{P_{\text{inc}}} = \left|\frac{E_{3,s}}{E_1}\right|^2 = \frac{(1-r^2)a^2}{1-2ra\cos\phi_R + r^2 a^2}.$$

At the resonant point ($\phi_R = m2\pi$, $m$ is an integer), the buildup factor can be reduced to $B_{\text{res}} = \frac{(1-r^2)a^2}{(1-ra)^2}$. Furthermore, the non-resonant and resonant buildup factor is related by $B(\phi_R) = B_{\text{res}} \frac{1}{1+b(1-\cos\phi_R)}$. Here, the parameter $b = \frac{2ra}{(1-ra)^2}$. If the attenuation is negligible ($a = 1$) and self-coupling coefficient is near to unit, then $b \xrightarrow[a \to 1, r \to 1]{} \frac{1}{2} B_{\text{res}}^2$. The first order derivation of buildup factor is

$$\frac{d}{d\phi_R} B(\phi_R) = \frac{-(1-r^2)a^2 2ra \sin\phi_R}{(1-2ra\cos\phi_R + r^2 a^2)^2} \quad (F5)$$

If the microresonator works around the half-maximum of the peak, buildup factor is sensitive to the normalized frequency detuning. To find the point that is most sensitive to the detuning, solve equation $\frac{d}{d\phi}\left(\frac{d}{d\phi}B(\phi_R)\right) = 0$, we find

$$\phi_0 = \arccos\left(1 - \frac{1}{b}\right) \quad (F6)$$

In fact, $B(\phi_0) = \frac{1}{2} B_{\text{res}}$, the FWHM of the buildup factor peak is $2\phi_0$. At the half-width of the buildup factor peak, the change of buildup factor due to small phase shift is

$$B(\phi_0) - B(\phi_0 + \delta\phi) \cong \frac{\sqrt{2}}{4} B_{\text{res}} \sqrt{b} \delta\phi \xrightarrow[a \to 1, r \to 1]{} \frac{1}{4} B_{\text{res}}^2 \delta\phi =$$
$$\frac{1}{\pi^2} \mathcal{F}^2 \delta\phi,$$

where $\mathcal{F}$ is finesse, related to buildup factor with relation $\mathcal{F} = \frac{\pi}{2} B$ in an all-pass resonator [66].

## APPENDIX G: DYNAMIC ANALYSIS OF OPTICAL MICRORESONATOR

For an all-pass resonator, assuming the input field is quasi-monochromatic light field with constant amplitude. The time-dependent phase shift $\delta\phi(t)$ occurred intracavity will result in the variation of the circulating field and output field. Here, we assume the amplitude of phase shift is small, satisfied $|\delta\phi(t)| \ll 1$.

The time-dependent circulating field $E_4(t)$ evolves: 1.

Current input contribution due to part of $E_1(t)$ couples into the resonator via $\kappa$. 2. Previous circulation contribution due to the field $E_4(t - T_R)$ from one round-trip earlier, attenuated by $a$ and phase-shift by $\phi_R + \delta\phi(t)$. Combining the field relations and round-trip dynamics: $E_4(t) = i\kappa E_1(t) + rE_3(t)$ and $E_3(t) = ae^{i(\phi_R + \delta\phi(t))}E_4(t - T_R)$ (before the perturbation, $E_3(t) = ae^{i\phi_R}E_4(t - T_R)$), we have the time-domain recursion:
$$E_4(t) = i\kappa E_1 + xe^{i\delta\phi(t)}E_4(t - T_R).$$
This can be rewritten as
$$E_4(t) = i\kappa E_1 + xE_4(t - T_R) + (e^{i\delta\phi(t)} - 1)xE_4(t - T_R).$$
Separating the $E_4(t)$ to steady-state part and perturbation part, $E_4(t) = E_{4,s} + \delta E_4(t)$, where $\delta E_4(t)$ is the small deviation due to $\delta\phi(t)$, thus
$$\underbrace{E_{4,s} + \delta E_4(t)}_{E_4(t)} = i\kappa E_1 + x\underbrace{\left(E_{4,s} + \delta E_4(t - T_R)\right)}_{E_4(t - T_R)} + (e^{i\delta\phi(t)} - 1)xE_4(t - T_R).$$
Because $E_{4,s} = i\kappa E_1 + xE_{4,s}$, after canceling the common $E_{4,s}$ on both sides, we get
$$\delta E_4(t) = x\delta E_4(t - T_R) + (e^{i\delta\phi(t)} - 1)xE_4(t - T_R).$$
Since $\delta E_4(t)$ is small, $E_4(t - T_R)$ can be replaced by its zeroth-order value $E_{4,s}$, $E_4(t - T_R) \approx E_{4,s}$, thus, the recursion is
$$\delta E_4(t) = x\delta E_4(t - T_R) + (e^{i\delta\phi(t)} - 1)xE_{4,s}.$$
The solution of this closed-form series can be found by unfolding that recurrence for $t$ after the perturbation starts,
$$\delta E_4(t - T_R) = x\delta E_4(t - 2T_R) + (e^{i\delta\phi(t - T_R)} - 1)xE_{4,s}.$$
$$\delta E_4(t - 2T_R) = x\delta E_4(t - 3T_R) + (e^{i\delta\phi(t - 2T_R)} - 1)xE_{4,s}.$$
...
Combined all recurrence relationships,
$$\delta E_4(t) = E_{4,s}\sum_{n=0}^{\infty} x^{n+1} \cdot (e^{i\delta\phi(t - nT_R)} - 1).$$
Hence the total circulating fields are
$$E_4(t) = E_{4,s} + E_{4,s}\sum_{n=0}^{\infty} x^{n+1} \cdot (e^{i\delta\phi(t - nT_R)} - 1) \quad \text{(G1)}$$

Similarly, the output field $E_2(t)$ evolves: 1. Current input contribution due to part of $E_1(t)$ output via reflection. 2. Circulation contribution due to the field $E_3(t)$ couples into the resonator via $\kappa$. Combining the field relations and round-trip dynamics: $E_2(t) = rE_1(t) + i\kappa E_3(t)$ and $E_3(t) = ae^{i(\phi_R + \delta\phi(t))}E_4(t - T_R)$. The output fields are
$$E_2(t) = rE_1(t) + i\kappa ae^{i(\phi_R + \delta\phi(t))}E_4(t - T_R)$$
Replace $E_2(t) = E_{2,s} + \delta E_2(t)$, we have
$$E_{2,s} + \delta E_2(t) = rE_1(t) + i\kappa ae^{i\phi_R}\left(E_{4,s} + \delta E_4(t - T_R)\right) + i\kappa ae^{i\phi_R}(e^{i\delta\phi(t)} - 1)E_4(t - T_R).$$
Because $E_{2,s} = rE_1 + i\kappa ae^{i\phi_R}E_{4,s}$, after canceling the common terms on both sides, we get $\delta E_2(t) = i\kappa ae^{i\phi_R}\delta E_4(t - T_R) + i\kappa ae^{i\phi_R}(e^{i\delta\phi(t)} - 1)E_{4,s} = i\frac{\kappa}{r}\delta E_4(t)$.

Thus, the transient respond of output field is
$$E_2(t) = E_{2,s} + i\frac{\kappa}{r}E_{4,s}\sum_{n=0}^{\infty} x^{n+1} \cdot (e^{i\delta\phi(t - nT_R)} - 1). \quad \text{(G2)}$$

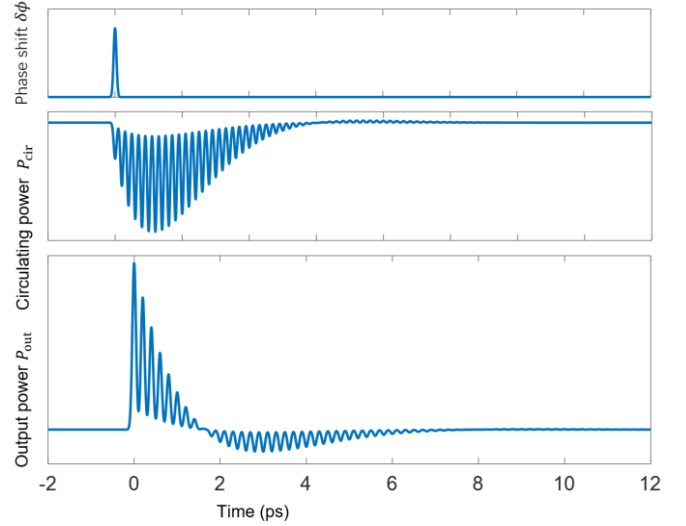

FIG. 5. Phase shift-induced dynamic of circulating power and output power for phase shift duration shorter than one round trip time. The setting of FWHM of Gaussian-shape phase shift is 0.1ps, one round trip time is 0.2ps, cavity lifetime is 5ps.

**APPENDIX H: THE NET ENERGY OUTPUT**

In the all-pass microresonator, assuming the input power is in steady state, the time-dependent phase-shift on the circulating field results in the power output
$$P_{out}(t) = P_{inc}\left|\frac{E_2(t)}{E_1}\right|^2.$$
Taking Eq. (G2) into above equation, after some calculations, the net power output can be further written as
$$P_{out}(t) \approx P_{out,s} + P_{inc}\frac{\kappa^2}{|1-x|^2}2\text{Re}\left[\left(\frac{1}{r^2}x^* - 1\right)\frac{\delta E_4(t)}{E_{4,s}}\right].$$
where $P_{out,s} = P_{inc}\left|\frac{E_{2,s}}{E_1}\right|^2$ is the steady state power output. Net energy output due to the small phase shift in the cavity can be calculated by
$$E_{net} = \int_{-\infty}^{+\infty} dt\, (P_{out}(t) - P_{out,s})$$
$$\approx 2P_{inc}\frac{\kappa^2}{|1-x|^2}\int_{-\infty}^{+\infty} dt\, \text{Re}\left[\left(\frac{1}{r^2}x^* - 1\right)\frac{\delta E_4(t)}{E_{4,s}}\right] \quad \text{(H1)}$$
Assuming the phase shift is last from $t = 0$ to $t = T_{int}$ and the phase shift amplitude is a constant,

$$\delta\phi(t) = \begin{cases} 0 & \text{for } t < 0 \text{ and } t \geq T_{\text{int}} \\ \delta\phi & \text{for } 0 \leq t < T_{\text{int}} \end{cases}.$$

If the duration of the phase shifts is no more than one round trip time, i.e., $T_{\text{int}} \leq T_{\text{R}}$, then

$$\frac{\delta E_4(t)}{E_{4,s}} = \begin{cases} 0 & \text{for } t < T_{\text{R}} \\ x^n(e^{i\delta\phi} - 1) & \text{for } nT_{\text{R}} \leq t < nT_{\text{R}} + T_{\text{int}} \\ 0 & \text{for } nT_{\text{R}} + T_{\text{int}} \leq t < (n+1)T_{\text{R}} \end{cases},$$

where $n = \lceil t/T_{\text{R}} \rceil$. The integral of real part in Eq. (H1) can be transformed into

$$\int_{-\infty}^{+\infty} dt\, \text{Re}\left[\left(\frac{1}{r^2}x^* - 1\right)\frac{\delta E_4(t)}{E_{4,s}}\right].$$

$$= \sum_{n=1}^{\infty} \int_{nT_{\text{R}}}^{(n+1)T_{\text{R}}} dt\, \text{Re}\left[\left(\frac{1}{r^2}x^* - 1\right)x^n(e^{i\delta\phi} - 1)\right].$$

$$= \sum_{n=1}^{\infty} T_{\text{int}} \text{Re}\left[\left(\frac{1}{r^2}x^* - 1\right)x^n(e^{i\delta\phi} - 1)\right].$$

$$= T_{\text{int}} \text{Re}\left[(e^{i\delta\phi} - 1)\frac{a^2 - x}{1 - x}\right].$$

$$\approx \frac{T_{\text{int}}\delta\phi}{|1-x|^2} ra(1 - a^2) \sin\phi_{\text{R}}.$$

Take above result into Eq. (H1), we have

$$E_{\text{net}} \approx P_{\text{inc}} 2T_{\text{int}}\delta\phi \frac{ra(1-r^2)(1-a^2)\sin\phi_{\text{R}}}{(1-2ra\cos\phi_{\text{R}} + r^2a^2)^2}.$$

Here, some factors can be replaced by the first order derivation of intracavity buildup factor shown in Eq. (F5). Thus, the net energy output can be simplified to

$$E_{\text{net}} = -P_{\text{inc}} T_{\text{int}}\delta\phi \frac{dB(\phi_{\text{R}})}{d\phi_{\text{R}}}(1-a^2)\frac{1}{a^2} \quad \text{(H2)}$$

This expression shows that net output energy depends on two factors: 1. The phase sensitivity of intracavity buildup factor $\frac{dB(\phi_{\text{R}})}{d\phi_{\text{R}}}$. 2. The intensity loss of every round trip $(1 - a^2)$. For no loss resonator, $E_{\text{net}} = 0$, it is impossible to detect the net energy output.

If the detuning phase is set at half-max point, $\phi_{\text{R}} = \phi_0$, then

$$E_{\text{net}} = -P_{\text{inc}} T_{\text{int}}\delta\phi \frac{(1-r^2)(1-a^2)\sqrt{ra}}{2(1-ra)^3}.$$

The finesse of all-pass resonator can be express as $\mathcal{F} = \frac{\text{FSR}_{\phi_{\text{R}}}}{\text{FWHM}_{\phi_{\text{R}}}} = \frac{\pi\sqrt{ra}}{1-ra}$. The circulating light intensity $I_{\text{cir}} = I_{\text{inc}}\frac{2\mathcal{F}}{\pi}$. For high-finesse microresonator ($a \to 1, r \to 1$), $\frac{(1-r^2)(1-a^2)}{(1-ra)^2} \approx 2$. Taking those expressions to above equation, then we have

$$E_{\text{net}} \xrightarrow[a \to 1, r \to 1]{} -\frac{1}{2} P_{\text{cir}} T_{\text{int}}\delta\phi.$$

Thus, the output field is proportional to the circulating light intensity.